\documentclass[]{article}
\usepackage{amsmath,amssymb}
\usepackage{graphicx}
\usepackage{enumerate}
\usepackage{natbib}
\usepackage{times}
\usepackage{bm}
\usepackage{url} 
\usepackage{tikz}
\usetikzlibrary{arrows, decorations.markings, matrix}
\usetikzlibrary{bayesnet}
\usepackage{setspace}
\usepackage{float}
\usepackage{xcolor}
\usepackage{epstopdf}

\title{Evaluating the relative contribution of data sources in a Bayesian analysis with the application of estimating the size of hard to reach populations}

\author{Jacob Parsons$^1$, Xiaoyue Niu$^1$, Le Bao$^1$\\
Department of Statistics, Penn State University, University Park, PA, USA$^1$\\
email: \texttt{lebao@psu.edu} }

\begin{document}

\maketitle

\begin{abstract}
When using multiple data sources in an analysis, it is important to understand the influence of each data source on the analysis and the consistency of the data sources with each other and the model. We suggest the use of a retrospective value of information framework in order to address such concerns. Value of information methods can be computationally difficult. We illustrate the use of computational methods that allow these methods to be applied even in relatively complicated settings.

In illustrating the proposed methods, we focus on an application in estimating the size of hard to reach populations. Specifically, we consider estimating the number of injection drug users in Ukraine by combining all available data sources spanning over half a decade and numerous sub-national areas in the Ukraine. This application is of interest to public health researchers as this hard to reach population that plays a large role in the spread of HIV. We apply a Bayesian hierarchical model and evaluate the contribution of each data source in terms of absolute influence, expected influence, and level of surprise. Finally we apply value of information methods to inform suggestions on future data collection.
\end{abstract}

\section{Introduction}

When performing an analysis using multiple data sources, it is important to understand each data source influences any resulting decisions, estimates, or predictions. As we shall illustrate, exploring the influence and potential influence of each data source allows one to assess the consistency of data sources, detect outlying portions of the data, identify aspects of a model that do not fit well with the available data, and help plan future research.  In what follows we make use of a value of information framework to provide a unified approach to measuring influence that addresses each of these goals.

The value of information was conceived of as a powerful tool for choosing between potential research regimes and planning experiments by measuring the expected improvement that additional information would provide \citep{raiffa1961} . For a period, the computational difficulty of computing certain key quantities prevented value of information methods from gaining widespread use. As a result, the focus of most recent research is on overcoming the computational difficulty of calculating quantities that are central to applying value of information methods \citep{ades2004,strong2015,keisler2014}. Previously, we have extended the scope of value of information method by providing a framework for evaluating the influence of different portions of a data set in Bayesian modeling setting \citep{Retrospect}. When applying value of information methods to influence analysis, the familiar computational issues take a new form which precludes the use of existing methods in most cases. One goal of the current work is to address this concern.

We are also interested in applying value of information methods to evaluate influence of different data sources to the problem of estimating the size of hard to reach populations.  Approaches to estimating population sizes that rely on different data sources result in very different estimates. New methods that combine multiple data sources are the most promising path to resolving this difficulty \citep{Okal2013,Abdul2015}. Previously, we have proposed a Bayesian hierarchical model for this purpose and applied the model to estimating the number of injection drug users in Ukraine, a hard to reach population that plays a large role in the spread of HIV \citep{Ukraine}.

We will begin by introducing the data and model that will serve as the setting and backdrop for analysis and exploration of the value of information. Next, we will review the value of information framework while expanding its scope beyond the standard presentation. After establishing the general framework, we will present an improved approach to computing the essential quantities and illustrate it with an example from the size estimation analysis. We will then be ready to apply the value of information framework to the Ukraine size estimation model and data. We present the analysis in three main phases: exploring the influence that each kind of data source has had on the fit of the model, exploring the influence that the data coming from particular locations has had, and finally providing recommendations for future data collection.

\section{The Value of Information}
\subsection{Prospective Value of Sample Information}
A typical development of value of information concepts will begin with the definitions and interpretations for the main measures. One of the central quantities used in this framework is the expected value of sample information \citep{raiffa1961}:
$$
EVSI(Y) =  E_{\theta }[L(a_0, \theta) - L(a_{Y} , \theta) ]
$$
Here $Y$ is a vector of observations that corresponds to an experiment or other source of information that we have not yet observed. $\theta$ is a parameter that determines the distribution of $Y$ and has a prior distribution summarizing the previous knowledge about the parameter. The goal is to choose an action $a$ from a set of possible actions $\mathcal{A}$ where a loss function $L(a, \theta)$ measures the quality of each action (note that estimation is a special case of this framework). Given only prior information the ideal action is
$$
a_{0} =  \arg \min \limits_{a \in \mathcal{A}} E_{\theta  }[L(a, \theta) ].
$$ 
If one observes $Y$ before making a decision, the chosen action would be
$$
a_{Y} =  \arg \min \limits_{a \in \mathcal{A}} E_{\theta \vert Y }[L(a, \theta) \vert Y].
$$ 
The expected value of sample information is then interpreted as the expected decrease in loss that would occur if the decision maker observes $Y$ rather than making the decision based only on prior information. There are limitations to this approach. For instance, the standard approach assumes the decision maker will be using the same prior and model when planning for data analysis and when determining the value of the information during the planning stage of an experiment. This disallows the use of priors of different strength levels during the planning and analysis stages of an experiment as is often employed using other approaches (see, for example, \cite{wang2002}). Another limitation is that this standard treatment of the value of information only applies when a Bayes decision rule is used. This disallows the use of this framework when using frequentist methods or even a risk averse Bayesian decision rule.

In what follows we shall refer to this expected value of sample information as the prospective expected value of sample information. For instance, if one is interested in the influence of a portion of the data $Y_{new}$ after we already observed $Y_{old}$, the retrospective expected value of sample information is given by 
$$
PVSI(Y_{new} \vert Y_{old} ) =  E_{\theta \vert Y_{old} }\Big[L\Big(a_{old}(Y_{old}), \theta\Big) - L\Big(a_{all} (Y_{old}, Y_{new})  , \theta \Big)  \vert Y_{old}\Big].
$$
We describe this quantity as ``prospective'' since it is the expected influence in terms of reduction in loss that would be provided by data that we have not yet observed. This distinction will be useful to differentiate this expectation from others that condition on a different set of data. We shall now consider measuring the influence of data that has already been observed.

\subsection{Retrospective Value of Information} 

Using the same notation as in the previous section, we define the retrospective expected value of sample information as
$$
RVSI(Y_{new} \vert Y_{old} ) =  E_{\theta \vert Y_{old}, Y_{new} }\Big[L\Big(a_{old}(Y_{old}), \theta\Big) - L\Big(a_{all} (Y_{old}, Y_{new})  , \theta \Big)  \vert Y_{old}, Y_{new} \Big].
$$

The retrospective expected value of sample information measures how much the addition of $Y_{new}$ is expected to have improved a decision after we already observed this portion of the data. That is, it is a measure of how much we think a piece of already collected information has influenced the quality of our decision as measured by a reduction in loss. This quantity is equivalent to an influence measure proposed by \cite{kempthorne1986} \citep{Retrospect}. 

An application of the retrospective expected value of information begins by partitioning a data set $Y_{all}$ into $n$ pieces $Y_1 , \dots, Y_n$. The next step is to compute $RVSI(Y_i \vert Y_{-i})$ for $i = 1, \dots, n$ where $Y_{-i}$ is $Y_{all}$ with the observations in $Y_{i}$ removed.  In this setting $RVSI(Y_i \vert Y_{-i})$ measures the reduction in loss by excluding only that portion of the data. For this to be meaningful, it is important to choose an appropriate loss function. For instance, in an estimation or prediction setting, the loss function should be a distance measure or error function. Comparing $RVSI(Y_i \vert Y_{-i})$ across values of $i$ is appropriate as the scale is the same in each case if the loss function is not changed. So, the portion of the data $Y_i$ with the highest retrospective value has had the largest influence on a decision.  

A portion of the data being more influential than other portions of the data does not necessarily suggest that the a portion of the data deviates from a model more than would be expected. For instance, in a linear regression setting a point with an extremely high leverage may have a larger influence on the fit of a line than other points even if it is fairly close to the line fit using the remaining data points. In this case the large influence of the point is expected even before we look at the actual value of the response variable. In the value of information setting $PVSI(Y_i \vert Y_{-i})$ plays a role similar to one use of leverage in the linear regression setting in that it provides a measure of how strongly we expect $Y_i$ to influence a decision before even observing the response. In comparison $RVSI(Y_i \vert Y_{-i})$ plays a role more similar to Cook's distance in a regression setting, measuring the influence of already observed quantities \citep{Retrospect}. 

A third quantity, the expected value of information ratio, provides a measure of how surprising a portion of the data is by comparing the prospective and retrospective value of information. We define this quantity as
$$
EVOIR(Y_i \vert Y_{-i}) =\frac{RVSI(Y_i \vert Y_{-i})}{PVSI(Y_i \vert Y_{-i})}.
$$
Note that by construction $EVOIR(Y_i \vert Y_{-i})$ has a mean of $1$ after taking the expectation over $Y_i|Y_{-i}$ because of the law of total expectation \citep{Retrospect}. Thus a crude interpretation of the expected value of information ratio is that a portion of the data is more influential than expected based on the rest of the data if it is greater than $1$ and less influential than expected if it is less than one. This quantity can be used to identify especially surprising portions of the data as such portions have an expected value of information ratio that is much higher than 1.

Once unusual points have been identified, further investigation should take place in order to understand the exact nature of the influence of these points. There are essentially four cases to consider when measuring influence using the retrospective value of information to measure influence.

First, the portion of the data under consideration has a low retrospective value and low expected value of information ratio, in which case including that data results in a decision that is similar to what would be done without that data. In this case further investigation is unlikely to yield any interesting results, unless there are concerns about data fabrication or a large percentage of the data portions under consideration have a very small expected value of information ratio. The latter situation may indicate an overly conservative model for example, due to a tendency to over estimate a variance parameter) or be indicative of a model neglecting real relationships between portions of the data.  

A second possibility is that a portion has high retrospective value but does not have a high expected value of information ratio because its prospective value is also high. In this case the portion of the data under consideration is expected to have a large influence due to the structure of the data and is also observed to have a large influence. This may be uninteresting, for instance, if the high prospective value is a result of being a larger portion of the data than the others under consideration. In such a situation, it may be worth breaking this portion of the data down further.  This situation may be more interesting if the higher prospective value is due to something more meaningful, for instance having a drastically different set of independent variables compared to the rest of the data. In such a case, it may be interesting to consider if the model is appropriately applied to this unusual class of data.   

A third possibility is that the portion of the data under consideration does not have a high retrospective value but has a high expected value of information ratio. In this case, some level of inconsistency with the model as fitted by the remaining data may be indicated. Further investigation should be done in this case to make sure that there is no transcription error when recording this data, that the data doesn't violate model assumptions, and an attempt should be made to identify ways in which this portion of the data differs from the remaining data. Regardless, the portion of the data does not have a substantial influence on the decision. 

The final possibility is that there is a high expected value of information ratio. This is the most worrisome possibility and one should investigate further as indicated in the previous combination. The main difference from the previous case is that conclusions based on the analysis are drawn into question since a portion of the data that is inconsistent with the remaining data has a large impact on the endpoints of the analysis.	

\section{Computation}
\subsection{Computation for a Quadratic Loss Function}
The most computationally intensive aspect of applying value of information methods lies in computing the prospective expected value of sample information. This is especially true when determining an appropriate action is computationally intensive. In this section we will first consider computing the expected value of information when an action is determined by minimizing the expected loss under a quadratic loss function. This corresponds to the very common practice in Bayesian analysis of using the posterior mean as an estimate or prediction. We will then adapt the resulting approach to other loss functions.

Let $Y_{old}$ and $Y_{new}$ be the data under consideration and $L$ be the loss function. If $L$ is a quadratic loss function, then
$$
a_{Y_{old}} = E(\theta \vert Y_{old}),
$$
$$
a_{Y_{old},Y_{new}} = E( \theta \vert Y_{old}, Y_{new}).
$$
\cite{Retrospect} show that for a quadratic loss function the retrospective value of sample information has the form
$$
RVSI(Y_{new} \vert Y_{old}) = (E[\theta \vert Y_{new} , Y_{old}] - E[\theta \vert Y_{old}] )^2,
$$
which is essentially how far the estimate moved in the Euclidean space by including $Y_{new}$ in the analysis in addition to $Y_{old}$. In this situation, the prospective expected value of sample information for the portion of the data $Y_{new}$ is
$$
PVSI(Y_{new} \vert Y_{old}) = E \big [  (E[\theta \vert Y_{new} , Y_{old}] - E[\theta \vert Y_{old}] )^2 \big\vert Y_{old} \big ].
$$
For a fixed $Y_{old}$ and $Y_{new}$ the retrospective value of sample information can typically be estimated directly from Monte Carlo draws from the posterior distribution conditional on just $Y_{old}$ and draws from the posterior conditional on the complete data composed of both $Y_{old}$ and $Y_{new}$. The prospective value is more difficult to compute as only $Y_{old}$ is bound by the expectation.  In this case $a_{Y_{old}}$ can be computed from a Monte Carlo sample directly but $a_{Y_{old}, Y_{new}}$ would be different at each draw in the Monte Carlo procedure. It would be inefficient to draw another Monte Carlo sample approximating the distribution of $\theta \vert Y_{old}, Y_{new}$ at each iteration in the original Monte Carlo sample that approximates the distribution of $Y_{new} \vert Y_{old}$  in order to determine the value of $a_{Y_{old}, Y_{new}}$ at each iteration. \cite{strong2015} present an approach to calculating the expected value of sample information when there are only a finite number of potential actions that avoids having to draw inner Monte Carlo samples from the distribution of $\theta \vert Y_{old}, Y_{new}$ at each iteration by replacing the inner samples with an estimate from a regression. The described method can be adapted to the case of a quadratic loss function using the following procedure:

\begin{itemize}
	\item Draw $\theta^{(i)}$ according to distribution of $\theta \vert Y_{old}$ for $i = 1,..., N$
	\item Draw $Y_{new}^{(i)}$ from the distribution of $Y_{new}^{(i)} \vert \theta^{(i)}, Y_{old}$ for $i = 1,..., N$ resulting in pairs $(Y_{new}^{(i)}, \theta^{(i)})$ drawn from the joint distribution of $Y_{new}, \theta \vert Y_{old}$
	\item Apply a non-parametric regression technique using the $Y^{(i)}_{new}$ as predictors and the $\theta^{(i)}$ as the response to approximate $E(\theta \vert Y_{old}, Y_{new})$ as a function of $Y_{new}$ by the fitted regression function $\hat{E}(\theta \vert Y_{old}, Y_{new})$.
	\item The prospective value of sample information is approximated by: 
	$$
	\frac{1}{N}\sum \limits_{i = 1}^N \big( \hat{E}(\theta \vert Y_{old}, Y_{new}^{(i)}) - E(\theta\vert Y_{old}) \big )^2
	$$
\end{itemize} 
 
This method is often orders of magnitude faster than the naive approach of using an inner Monte Carlo procedure for each draw of an outer Monte Carlo assuming an appropriate regression approach can be found. The above scheme adapting the method of \cite{strong2015} can be improved further in many cases. To this end we propose a modification of this procedure. Let $\theta_T$ be the target parameter and suppose $\theta_N$ are a set of nuisance parameters. Let $\theta = (\theta_T, \theta_N)$ be the vector containing all the parameters of the model.

\begin{itemize}
	\item Draw $\theta^{(i)} = ( \theta^{(i)}_T, \theta^{(i)}_N)$ according to distribution of $= ( \theta_T, \theta_N) \vert Y_{old}$ for $i = 1,..., N$
	\item Draw $Y_{new}^{(i)}$ from the predictive distribution of $Y_{new}  \vert \theta, Y_{old}$ for $i = 1,..., N$. We now have a draw from  $Y_{new},\theta_T, \theta_N \vert Y_{old}$
	\item Compute $Z^{(i)} = E(\theta_T \vert \theta_N^{(i)}, Y_{old}, Y_{new}^{(i)})$ for $i = 1,..., N$.
	
	\item Apply a non-parametric regression technique using the $Y^{(i)}_{new}$ as predictors and $Z^{(i)}$ as the response to approximate $E(\theta_T \vert Y_{old}, Y_{new})$ as a function of $Y_{new}$ using the fitted regression function $\hat{Z}(Y_{new})$.\\
	
	\item The prospective value of sample information is approximated by: 
	$$
	\frac{1}{N}\sum \limits_{i = 1}^N \big( \hat{Z}( Y_{new}^{(i)}) - E(\theta_T\vert Y_{old}) \big )^2
	$$

\end{itemize}

The above procedure assumes  $E(\theta_T \vert \theta_N^{(i)}, Y_{old}, Y_{new}^{(i)})$ can be determined quickly at each step. This is often the case (for example, whenever a Gibbs update is available). The reason that this procedure is an improvement over the original scheme is that $Z = E(\theta_T \vert \theta_N, Y_{old}, Y_{new}^{(i)})$ is a random variable with having mean of $E(\theta \vert Y_{old}, Y_{new})$ but with a smaller conditional variance (on average) than $\theta \vert Y_{old}, Y_{new}$. This allows for the regression to be more efficient when there is a low signal to noise ratio. This revised procedure is especially useful in a retrospective value of information analysis, as $Y_{old}$ already typically provides a substantial amount of information about $\theta$ resulting in $Y_{new}$ having a smaller effect on any decisions made.  

The effectiveness of the above schemes are heavily dependent on the regression method applied. If the regression yields a bad approximation of the mean function, then the computed expected value of sample information is unreliable. It is therefore important to check that the fitted regression function is a good approximation of true conditional mean.

\subsection{Computation for Other Loss Functions}

So far, we have only considered the case of a quadratic loss function. This choice of loss function is convenient in that the conditional mean function will be the function that minimizes the expected quadratic loss and most regression techniques are designed to estimate the conditional mean of a response variable as a function of the set of predictors. Moreover, a simple way of reducing the noise in the regression is available in under a quadratic loss function.

Now we shall consider a generic loss function $L(a, \theta)$. The same general procedure will still hold as long as we can find a method for estimating the true Bayes decision function
$$
a_{Y_{new}} = \arg \min \limits_{a} E \big ( L(a, \theta) \vert Y_{new}, Y_{old} \big )
$$
based on a finite sample $(Y_{new}^{(1)}, \theta^{(1)}), \dots , (Y_{new}^{(N)}, \theta^{(N)})$ from the joint conditional distribution of $Y_{new}, \theta \vert Y_{old}$. Without specifying some kind of form for the Bayes decision function, expecting to solve this would be hopeless. Any chosen method for estimating $a_{Y_{new}}$ should be flexible enough that it can approximate a set of functions that could plausibly contain the true function while avoiding over-fitting to the data (resulting in an over estimate for the $PVSI(Y_{new} \vert Y_{old})$). Typically multiple approaches to estimating $a_{Y_{new}}$ should be attempted and compared. Comparing different methods of estimation can be done by comparing the average loss based on a sample from the distribution $Y_{new}, \theta \vert Y_{old}$ not used to generate estimated functions. 

Once a method of estimation has been chosen, the procedure is very similar to the first method outlined in the previous section:

\begin{itemize}
	\item Draw $\theta^{(i)}$ according to distribution of $\theta \vert Y_{old}$ for $i = 1,..., N$
	\item Draw $Y_{new}^{(i)}$ from the from the distribution of $Y_{new}^{(i)} \vert \theta^{(i)}, Y_{old}$ for $i = 1,..., N$ resulting in pairs $(Y_{new}^{(i)}, \theta^{(i)})$ drawn from the joint distribution of $Y_{new}, \theta \vert Y_{old}$
	\item Apply the estimation technique using the $Y^{(i)}_{new}$ as predictors and the $\theta^{(i)}$ as the response to approximate $a_{Y_{new}} = \arg \min \limits_{a} E \big ( L(a, \theta) \vert Y_{new}, Y_{old} \big )$ as a function of $Y_{new}$. Denote the estimated function as $\hat{a}_{Y_{new}}$.
	\item The prospective value of sample information is approximated by: 
	$$
	\frac{1}{N}\sum \limits_{i = 1}^N L(a_0 , \theta^{(i)}) - L(\hat{a}_{Y_{new}}, \theta^{(i)})
	$$
\end{itemize} 

One big difference in this case is that no general method of noise reduction is available.

\subsection{Artificial Neural Networks as an Estimation Technique}

When using the approach to calculating $PVSI(Y_{new} \vert Y_{old})$ outlined above, different regression techniques will have different restrictions on what loss functions can be used and what decision functions can be approximated. Luckily, there is an available estimation technique that has shown success in a large variety of settings involving a variety of loss functions and settings. Artificial neural networks are a flexible class of models that can approximate a large class of functions and for which many approaches to avoiding over-fitting are available. Furthermore, standard techniques for fitting such networks allow for a wide class of loss functions. There exists a vast literature on artificial neural networks and summarizing the current research in the field is beyond the scope of this paper. Several introductions to and overviews of the field are available \citep{ann1, ann2, ann3}. In what follows we will outline the specific class of fully connected neural networks used in the later application.

An artificial neural network can be conceived as consisting of multiple layers in sequence. The first layer, present in all such networks, is the input layer. In our application this layer will consist of a number of input nodes equal to the dimension of $Y_{new}$ that will serve as the argument to the estimated function. The values of the $i$th node in the input layer will be denoted by $X_{0p}$ and the vector of all values for the row will be $X_0$. At each iteration of the previously described Monte Carlo procedure $X_0 = Y_{new}^{(i)}$. There will also be $L$ hidden layers (typically, multiple values of $L$ are tried). The $l$th hidden layer will have $n_l$ nodes. The $p$th node of the $l$th hidden layer will be given by
$$
X_{lp} = \sigma( a_{lp}^t X_{(l - 1)}  + b_{lp}).
$$

Here the $n_{l-1}$ dimensional weights $a_{lp}$ and the bias terms $b_{lp}$ are the parameters that need to be estimated when fitting the network. The activation function $\sigma$ is taken to be a standard sigmoidal function here but other choices are possible. The last layer in all networks, the output layer, corresponds in our application to the output of the decision function, $\hat{a} (Y_{new})$. The $p$th node of the output layer is given by
$$
X_{(L + 1)p} = \sigma_{output} ( a_{(L + 1)p}^t X_{((L + 1) - 1)}  + b_{(L + 1)p} ).
$$

A different final activation function, $\sigma_{output}$, is used depending on the sort of action being considered. In the case that the response is a probability, for instance, a $logit$ function can be used and an identity function can be used in the case that the action space is the entire real line. Categorical variables can be encoded using dummy variables.

Fitting an artificial network can be done using the back-propagation algorithm, an application of Newton's method and the chain rule from calculus \citep{ann4}. It is more common to use more complicated procedures that are often more efficient and attempt to mitigate over-fitting. In our application, we apply the
Adadelta method, an iterative gradient descent optimization method that requires no manual setting of a learning rate, to fit the model \citep{ann5}. Dropout, the practice of leaving out a portion of the nodes at each iteration of the gradient descent, was used as a regularization technique to avoid over-fitting \citep{ann6}.

Every time that an expected value of sample information is estimated in the following sections, we use an artificial neural network as described above considering between 1 and 3 hidden layers and dropout rates between $.3$ and $.7$. These hyper parameters are selected by comparing performance on a sample drawn from the predictive distribution of $Y_{new}  \vert \theta, Y_{old}$ that were not used to fit the networks using the expected loss as a comparison metric. 

\section{Results for Prevalence of Injection Drug Use in Ukraine}
\label{sec:results}

In this section, we illustrate the use of the value of information framework discussed above in the setting of estimating the number of injection drug users in Ukraine. We begin by laying out the available data sources and the model used in section \ref{sec:model}. Next, in section \ref{sec:compExamp}, we provide a concrete example of computing the value of information measures and a comparison of multiple regression techniques in the computation of the value of information for a particular subset of the available data in the context of estimating the national prevalence of injection drug use in Ukraine in 2007. In section \ref{sec:prevInf}, we discuss the influence of each data source on estimating the prevalence of injection drug use on a national level. Then in sections \ref{sec:multInf} and \ref{sec:nsuInf} we discuss the influence of each data source on the estimates for the bias of each data source. We then transition from discussing the influence of each data source to considering the influence of each site from which data was collected in section \ref{sec:siteInf}. Finally, we use prospective value of information measures to provide guidance on the collection of additional data in section \ref{sec:futObs}.

\subsection{The Model}
\label{sec:model}

\begin{table}[]
	\caption {Number of cities for which multiplier estimates are available for each subgroup and year combination. The last column represents the year and number of cities that have the network scale-up estimates.} \label{tab:data} 
	\begin{tabular}{lllllllll}
		& DTF &DTP &Hospital& NGO &Prevention& SMT &Survey &Network Scale-up\\
		2007&  0  & 0  &     14 &  0  &        0 &  0  &    0 &0\\
		2008&  0  & 0  &     0  &  0  &        0 &  0  &    0 &27\\
		2009&  0  & 0  &      0 &  0  &        0 &  0  &   14 &0\\
		2010&  0  & 0  &     12 & 26  &       21 & 23  &    0 &0\\
		2013&  0  & 0  &      0 &  0  &        0 &  0  &   27 &0\\
		2014&  0  &27  &     27 &  0  &        0 & 24  &    0 &0\\
		2015& 27  &27  &     27 & 27  &        0 & 23  &    0 &0\\
		\\
	\end{tabular}
\end{table}

In Ukraine, two different methods have been applied to estimate the number of injection drug users in $27$ different cities: the multiplier method and the network scale-up method. The multiplier method uses two pieces of information to generate an estimate of the size of a population: an estimate for the proportion of the target population that fall into some subgroup of the target population ,$P$, and the true number of individuals in the subgroup, $Y$. In order to apply the multiplier method, we must have a subgroup with known size. In Ukraine, the subgroups of injection drug users that have been considered are injection drug users that stay in a drug treatment facility(DTF), those that participate in a drug treatment program (DTP), those that are hospitalized due to injection drug related causes (Hospital), those that undergo HIV rapid tests provided by non-government organizations (NGO), those that are registered to certain HIV prevention services (Prevention), those that undergo substitution maintenance therapy (SMT), and those that participated in previous behavioral surveys (Survey) \citep{Berlava2010, Berlava2012, Berlava2017}. The network scale-up method generates an estimate of a target population based on social networks reported in a general survey of the entire population. This method has only been applied during a single year in the Ukraine \citep{Paniotto2009}.

We previously presented a Bayesian hierarchical model for combining multiple data sources and used it to generate estimates for the number of injection drug users in Ukraine \citep{Ukraine}. We use essentially the same model here only replacing improper priors on beta distributions with proper priors. The results of the model are nearly identical.

The main estimands in the analysis were the prevalence of injection drug users, $\pi_{it}$ and the number of injection drug users $n_{it}$ for each $i$th city and $t$th year combination. These two quantities are linked in the model by
\begin{align}
n_{it} \sim \text{Binomial}(R_{it}, \pi_{it}).
\end{align}
where $R_{it}$ is the known total population size for the corresponding year and city combination. We link the prevalence at a site to the prevalence in the previous year:
\begin{align}
logit(\pi_{i(t+1)}) \sim N( logit(\pi_{it}) + \phi_{t}, \sigma_\pi^2 ).
\end{align}
$\sigma_\pi^2$ measures how weak the relationship is and $\phi_{t}$ is the national average change in proportion for a year on the logit scale. The initial year and trend terms for each year have the following distributions:
\begin{align}
\phi_{t} \sim N(0, \sigma_\phi^2), \nonumber\\
\pi_{i0} \sim Beta(\alpha_0, \beta_0).
\end{align}

We take a similar approach to linking the prevalence, $p_{ijt}$, of a subgroup $j$ among the injection drug using population of a given city and year combination to the number of individuals in that subgroup, $Y_{ijt}$:
\begin{align}
Y_{ijt} \sim \text{Binomial}(n_{it} , p_{ijt}).
\end{align}
As before, we link the proportion of drug users falling into a subgroup for each year in the following way:
\begin{align}
p_{ij0} &\sim Beta(\alpha_j, \beta_j),\nonumber\\
logit(p_{ij(t+1)}) &\sim N( logit(p_{ijt}) + \eta_{jt}, \sigma_p^2 ),\\
\eta_{jt} &\sim N(0, \sigma_\eta^2).\nonumber
\end{align}

Here, $\eta_{jt}$ is the average change across sites in the proportion of injection drug users falling into a particular subgroup in the year $t+1$ compared to the previous year $t$. $\sigma_p^2$ is a variance term that measures the variation of a subgroup between years. 

The estimate for the prevalence, $P_{ijt}$, of a specific subgroup $j$ for a year and city combination is not assumed to be unbiased:
\begin{align}
\text{logit}(P_{ijt}) = \text{logit}(p_{ijt}) + \theta + \delta_i + \gamma_j  + \epsilon_{ijt},
\end{align}
$\theta$ is the average bias on the logit scale of the estimates for the proportion of injection drug users falling into each of the subgroups across all sites and subgroups. $\theta + \delta_i$ is the average bias for the proportion estimates at a particular site. $\theta + \gamma_j$ is the average bias for the proportion estimates of a particular subgroup. The bias terms and error have the following distributions where $G_{ijt}$ is the sample size of the survey used to generate the corresponding estimate:
\begin{align}
\delta_i \sim N(0, \sigma^2_\delta),\nonumber\\
\gamma_j \sim N(0, \sigma^2_\gamma),\\
\epsilon_{ijt} \sim N(0, \sigma^2_\epsilon / G_{ijt})\nonumber.
\end{align}
$\sigma^2_\epsilon$ is the variance for a proportion estimate based on a sample size of $1$ on a logit scale. We model the network scale-up method estimates on a log scale and allow for these estimates to be biased. 
\begin{align}
log(N_{it}) \sim N[ log(n_{it}) + \mu, \frac{\tau S^2_{it}}{n_{it}^2} ].
\end{align}
Finally, we have the following prior distributions
\begin{align}
&\mu \sim N(0, 1) \nonumber\\
&\theta \sim N(0, 10) \nonumber\\
&\frac{\alpha_j}{\alpha_j + \beta_j} \sim Uniform(0,1) \\
&\alpha_j + \beta_j \sim Gamma(.01,.01) \nonumber\\
&\tau, \sigma^2_\pi, \sigma^2_p,  \sigma^2_\eta, \sigma^2_\phi, \sigma^2_N, \sigma^2_\epsilon \sim InverseGamma(.5,.5) \nonumber
\end{align}

In the remainder of this section we will explore the influence and consistency of the data corresponding to each multiplier subgroup and the portion of the data corresponding to the network scale-up method using a retrospective value of information analysis. We will then apply a prospective value of information analysis to evaluate the value of collecting new data for each data source and make recommendations about the application of size estimation methods in the future. 

For the retrospective analysis we will consider the following parameters: the bias of the multiplier method subgroup proportion estimates on the logit scale $\theta$, the bias of the network scale-up estimates on the log scale $\mu$, and the national average prevalence for each year. We shall use a quadratic loss function in each case: the loss function for $\theta$ is
$$
L(\hat{\theta},\theta) = (\hat{\theta} - \theta)^2.
$$
Similarly the loss function for $\mu$ is
$$
L(\hat{\mu},\mu) = (\hat{\mu} - \mu)^2.
$$
The loss function used for the national average prevalence for each year is given by
$$
L(\hat{l}_{t}, l_{t}) = (\hat{l}_{t} - l_{t}) )^2
$$
where $l_{t}$ is the national average logit prevalence for the year $t$. In particular, in the first year, the average prevalence on the log scale is given by
$$
l_{0} = digamma(\alpha_0) - digamma(\beta_0).
$$
Recall that the digamma function is the logarithmic derivative of the more well known gamma function. For $t>0$, the national average logit prevalence for the year $t$ can be expressed in recursive manner by taking into account the average drift in logit prevalance $\phi_t$:
$$
l_{t} = l_{t-1} + \phi_t.
$$
We also consider estimating the total loss in estimating the average logit prevalence using the loss function:
$$
L(\hat{l}, l) = \sum \limits_{t} (\hat{l}_{t} - l_{t})^2.
$$

\subsection{Computation Example: Influence of Hospitalization Data on Prevalence Estimates}
\label{sec:compExamp}

\begin{table}
	\centering

	\begin{tabular}{l|r|r|r}
		Model        & Mean Squared Error & Estimated PVSI  &Run Time  \\
		\hline
		ANN          & 0.067              & 0.61            &46 Min    \\
		GAM          & 0.103              & 0.69            &54 Min    \\
		Linear       & 0.309              & 0.47            &$<1$ Min          
	\end{tabular}

\caption {A comparison of four regression techniques used in the hospitalization example. Mean squared error is on the logit scale and is calculated using predictions for data not used to fit the model. All models were fit using the same hardware. ``Run time'' refers to the length of time required to fit the model.} 
	\label{tab:models} 
\end{table}

To illustrate the use of the procedure described in section 3.1 we consider calculating the prospective value of sample information for the hospitalization subgroup portion of the multiplier method data when estimating the average prevalence of injection drug use in Ukraine during 2007, $l_0$ using the squared error loss function
$$
L(\hat{l}_{0}, l_{0}) = (\hat{l}_{0} - l_{0})^2.
$$
The first step in the procedure is to sample all of the parameters from the posterior distribution. In this case a Metropolis-Hastings algorithm is used to draw 15000 samples after thinning from the posterior distribution conditional on all the data besides the hospitalization data. The second step is generating the missing multiplier data for the 94 site and year combinations for which hospitalization data exists in the full data set. 

The third step is where an element of choice enters into algorithm. In the finite decision space setting, \cite{strong2015} use a generalized additive model (GAM) for the non-linear regression technique in this step. In our calculations, we use an artificial neural network as the main regression model, but a comparison to a GAM and a linear model is shown in Table \ref{tab:models}.

The target quantity in this case is average logit national prevalence of sites during the first year. Combining the notations of section \ref{sec:model} and 3.2, the $i$th Monte Carlo draw for this quantity is
$$
 Z^{(i)} = l_0^{(i)}  = digamma(\alpha_0^{(i)}) - digamma(\beta_0^{(i)}).
$$
This quantity will serve as the response in the regression while the predictors will be the simulated observations for the hospitalization data sampled from the posterior predictive distribution conditional on the reduced data set. The fitted values from the regression serve as estimates for the  mean $l_0$ conditional on each simulated version of the hospitalization data and is used to yields an estimate for the prospective value of the hospitalization data.

The generated set of data is split into two groups using the Kennard-Stone algorithm: a training set used to fit the regression models and a test set used evaluate and compare the fit of the models \citep{kennard1969}. 

When comparing the models, the mean squared error is the ideal quantity to compare as the actual conditional mean is expected to minimize this criterion. It can be seen from Table \ref{tab:models} that the artificial neural network model had the smallest mean squared error and also took less time to fit than the generalized additive model. As would be expected, the linear model was much quicker to fit but performed much worse than the other two approaches. 

The time required to fit the artificial neural network includes the time taken to fit the model for multiple choice of tuning parameters. In particular multiple dropout rates are considered. The table shows the results for a two-layer neural network with 100 nodes in each hidden layer and a dropout rate of $.5$. Other choices of dropout rate also have a lower mean squared error than the GAM. For instance, a dropout rate of $.1$ for the same network structure results in a mean squared error of $.078$.

\subsection{Influence of Data Sources on Prevalence Estimates}
\label{sec:prevInf}

\begin{figure}
	\centering
	\includegraphics[height=8cm]{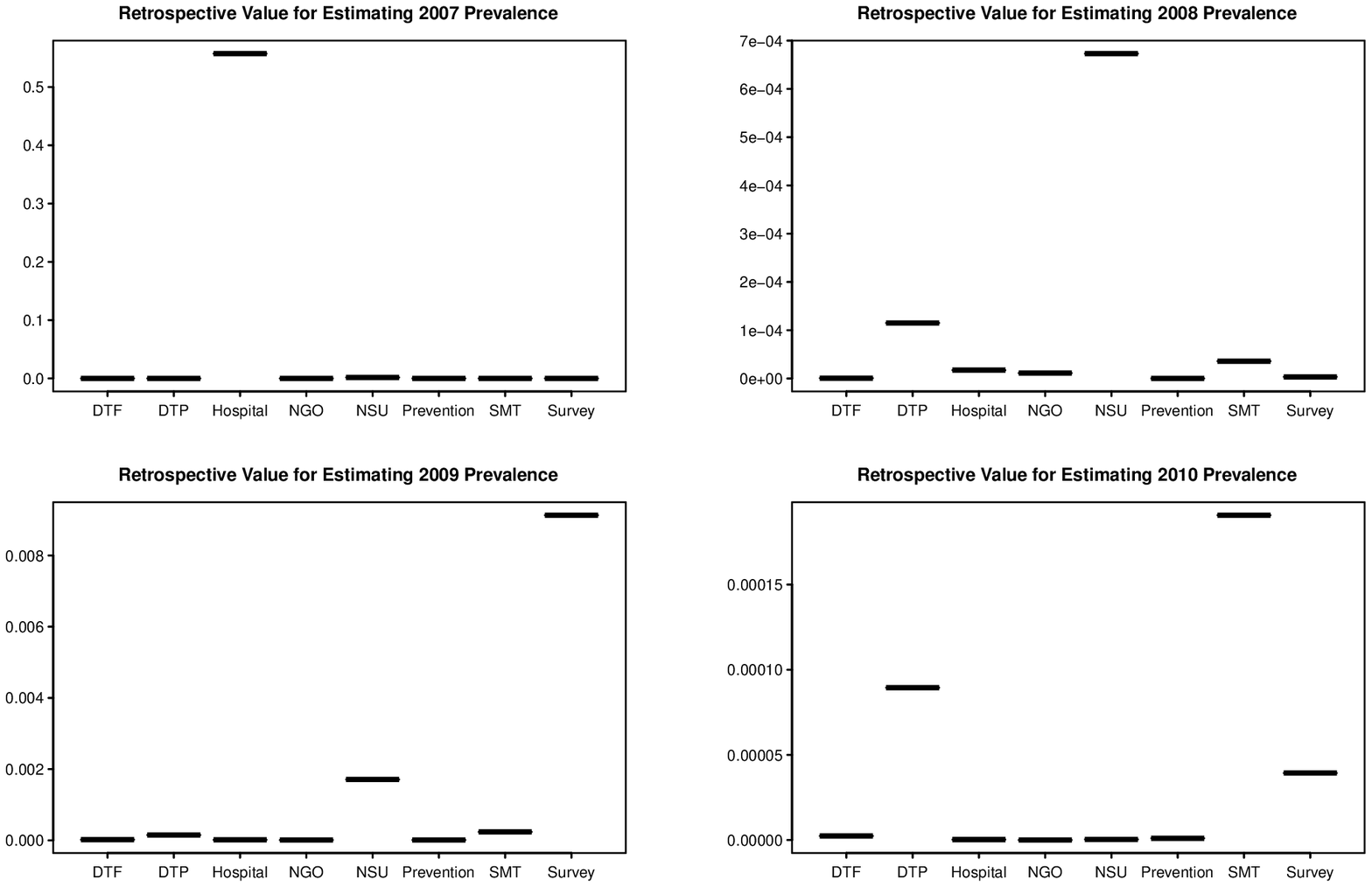}
	\includegraphics[height=8cm]{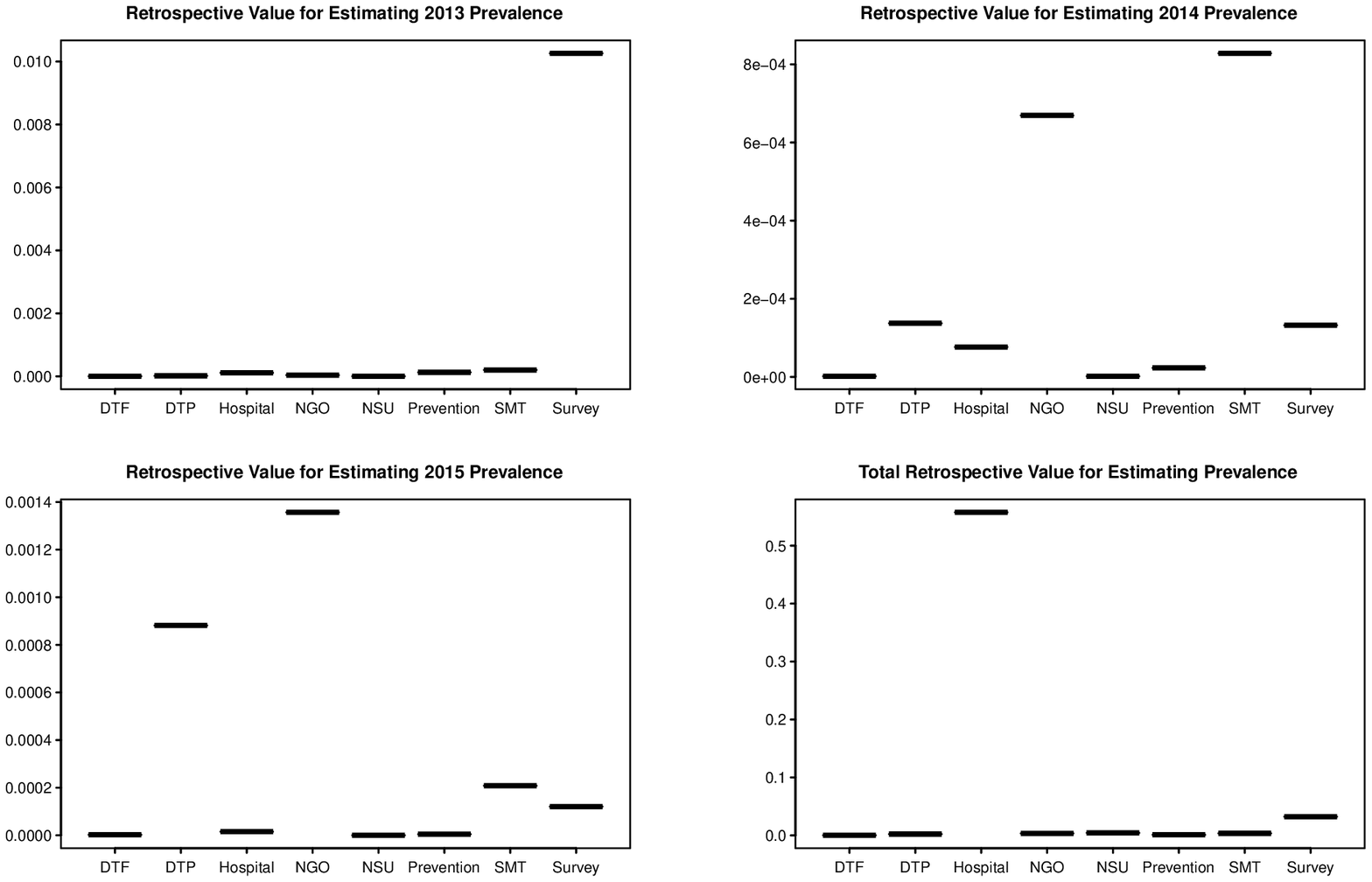}
	\caption{Retrospective value of sample information for estimating the prevalence of each year with data by data source. The total retrospective value is the the retrospective value of a data source for estimating the mean prevalence of all years considered as a vector.}
	\label{fig:RVSI}
\end{figure}

In an ideal situation, the estimated prevalence of a given year would be based on a variety of data sources as this would lead to more robust conclusions. As seen in Figure \ref{fig:RVSI} the prevalence estimates for all of the early years are dominated by a couple data sources. This is largely a result of the fact that for these only one or two data sources were present to inform an estimate.

\begin{figure}
	\centering
	\includegraphics[height=16cm]{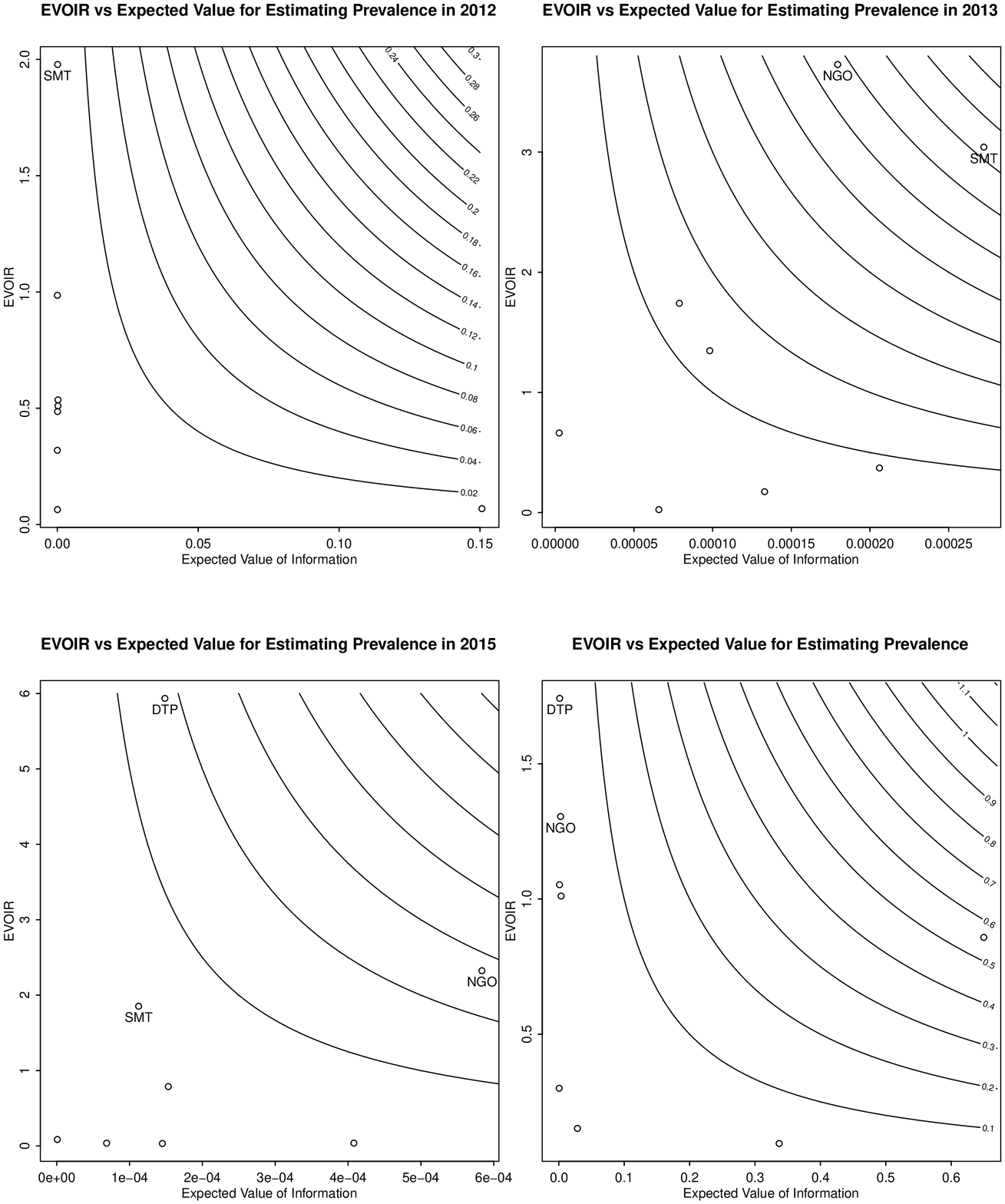}
	\caption{The expected value of information ratio plotted against the prospective expected value of information with contours for the retrospective value of information for the same partition of data.}
	\label{fig:EVOIR}
\end{figure}

\begin{figure}
	\centering
	\includegraphics[width=13cm]{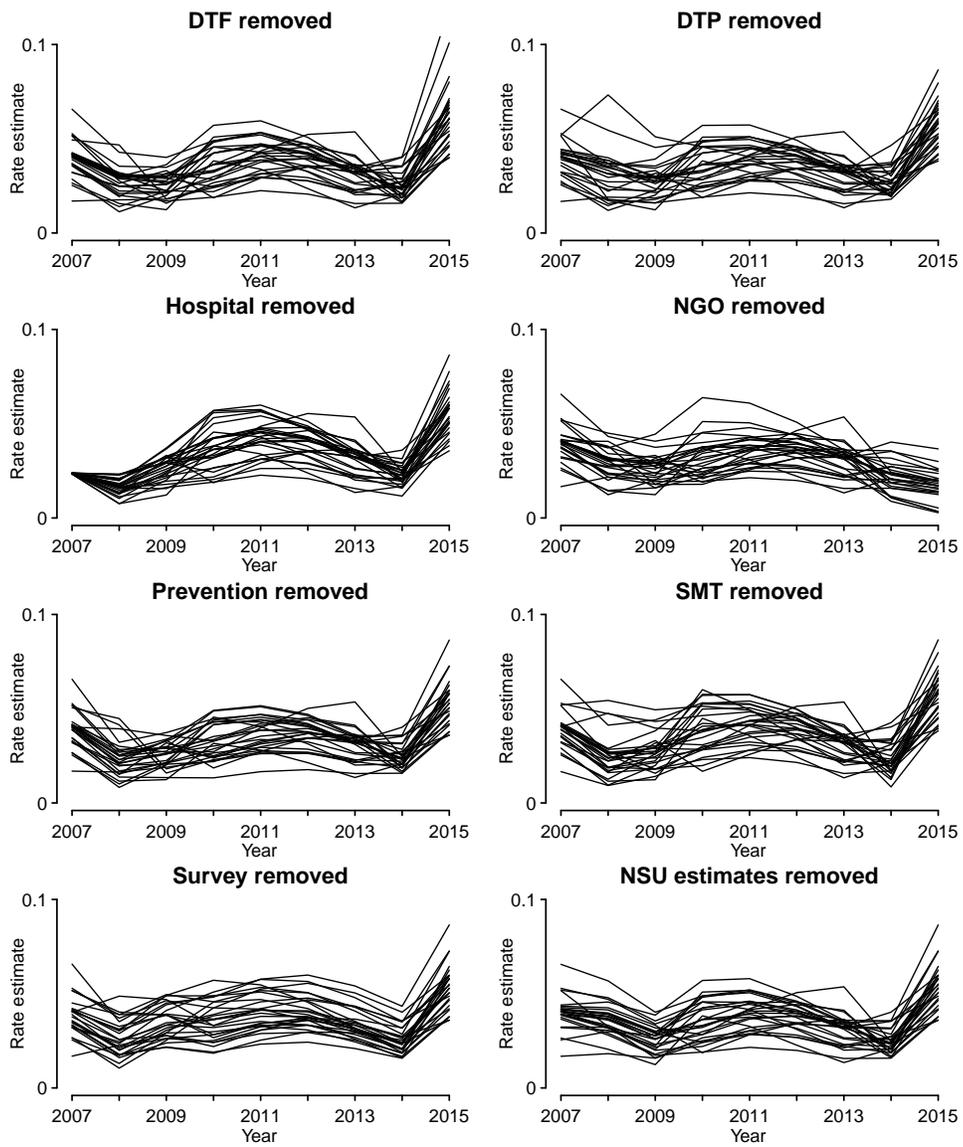}
	\caption{The estimated prevalence curves for each site estimated with each data source removed.}
	\label{fig:prevalence}
\end{figure}

More recently, there was an attempt to collect information on more subgroups in order to arrive at more robust conclusions. In these later years the influence of the NGO, SMT, and DTP subgroups stand out from the other groups in that they have a consistently surprising influence on the prevalence estimates as seen in Figure \ref{fig:EVOIR}. A subgroup had a high EVOIR because it suggested inconsistent results compared to other subgroups. We have already commented on the lack of stability of the proportion estimates not seen in the other sites for the NGO subgroup that likely accounts for the general level of surprise for this source. The SMT and DTP subgroup data, however, do not exhibit this behavior and have rather stable estimates across years. It is in this situation that it is important to consider the interactions of the level of influence that a portion of the data has. In particular, the SMT and DTP subgroups are no longer seen as especially surprising if the NGO subgroup is excluded from the analysis. Indeed, the NGO subgroup causes a sharp increase in the estimated prevalence in 2014 and 2015 that is not suggested by the remaining data as seen in Figure \ref{fig:prevalence}. In conclusion, the NGO subgroup appears to suggest a level of prevalence that is inconsistent with the other data sources while the other sources yield a fairly consistent set of estimates. As noted in the previous subsection, it may be more appropriate to treat the NGO subgroup as two different subgroups, one for the year 2010 and one for the year 2015.

It should be noted that the DTF and Prevention subgroups have nearly no impact the estimates for the prevalence of injection drug user in any year. This is largely a result of being present in only single year each resulting in them providing information about their own subgroup specific bias more than anything else. For this reason it is important to have multiple years of data for a subgroup. It is also for this reason that treating splitting the NGO subgroup into two subgroups would have a similar effect to removing it from the model altogether.  

\subsection{Influence of Data Sources on Proportion Estimate Bias}
\label{sec:multInf}

\begin{figure}
	\centering
	\includegraphics[height=6cm]{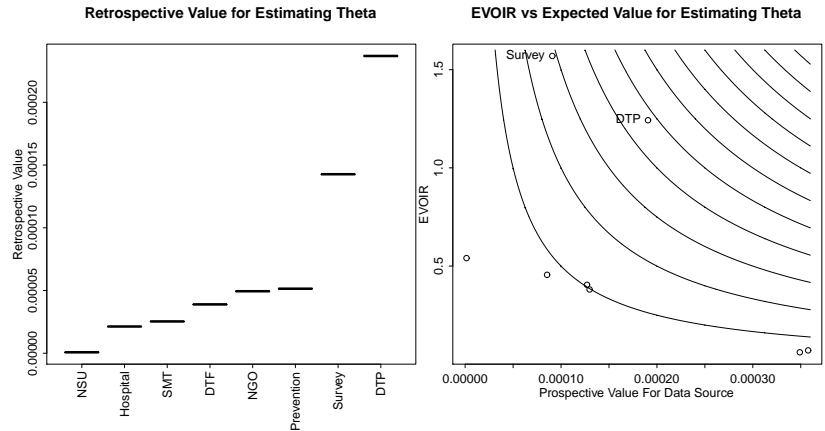}
	\caption{On the left is the retrospective value of information for each multiplier subgroup and the network scale-up data. The right plot shows the expected value of information ratio plotted against the prospective expected value of information with contours for the retrospective value of information for the same partition of data.}
	\label{fig:theta}
\end{figure}

We shall begin our exploration by examining the multiplier method subgroup proportion estimates bias term $\theta$. As can be seen from Figure \ref{fig:theta} two particular multiplier subgroups stand out as having had a much larger influence on the model fit than the other subgroups or network scale-up estimates: the Survey and DTP subgroups. 

In interpreting right hand plot in Figure \ref{fig:theta} it should be recalled that the expected value of information ratio on the vertical axis measures how surprising the observations are while the prospective expected value captures the level of influence due to structural aspects of the data: sample sizes, the number sites per year, etc. It can can be concluded that the larger level of influence for the Survey and DTP subgroups is largely due to how surprising the observations were compared to what is expected from the model fit according to the other sources of data rather than the structural aspects of the data. This is indicated by the fact that these two subgroups have a middling prospective expected value of sample information but the highest expected value of information ratios.

Despite having a higher expected value of information ratio than the other observations, these two subgroups do not have a worrisome level of surprise. According to the posterior predictive distribution conditional on the remaining data, we would expected the DTP subgroup to have a higher influence with a posterior probability $.19$ and the Survey subgroup to higher influence with posterior probability $.23$. The average expected value of information ratio among all the subgroups and network scale-up data $.59$ which is less than $1$ the theoretical average expected value of information ratio. We conclude that the multiplier data subgroups and the network scale-up data tend to give consistent estimates for the average bias of the multiplier subgroup proportion estimates. 

\subsection{Influence of Data Sources on Network Scale-Up Bias}
\label{sec:nsuInf}

\begin{figure}
	\centering
	\includegraphics[height=6.5cm]{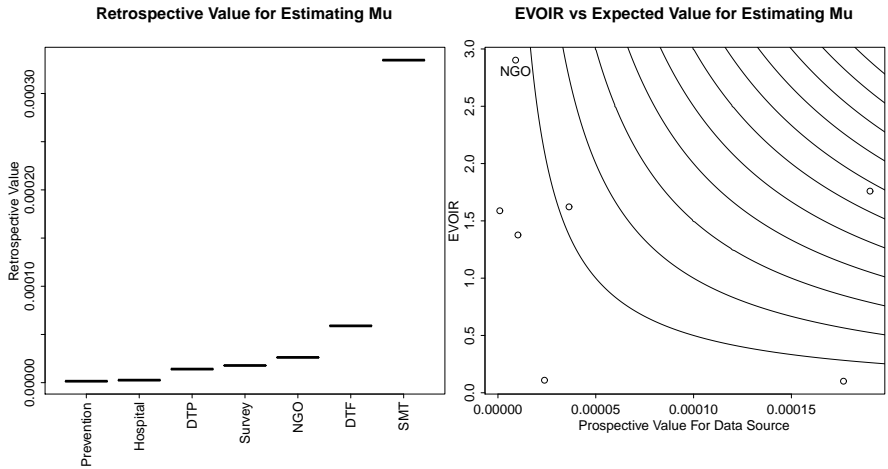}
	\caption{On the left is the retrospective value of information for each multiplier subgroup and the network scale-up data. The right plot shows the expected value of information ratio plotted against the prospective expected value of information with contours for the retrospective value of information for the same partition of data.}
	\label{fig:mu}
\end{figure}

We now move on to an examination of the influence of the data sources on the network scale-up bias on a log scale, $\mu$. It is important to note that the influence of a data source on the bias of another data source allows us to understand the consistency of the data sources. Thus, it is important to consider the impact of the multiplier method estimates on the fitted bias parameter of the network scale-up estimates. Again, we will use a squared error loss function. We should begin by discussing the influence that the network-scale up data itself had. The network scale up data has multiple orders of magnitude more prospective value on estimating the network scale-up bias than any of the multiplier subgroups (a prospective value of $.72$). This is because we are left with an untouched weak prior distribution over this bias if the network scale-up data is removed. So, comparing its influence to the other data sources is inappropriate in this context. Also because of the weak prior information, the expected value of information ratio is $.04$ which is very low compared to $1$. So, we will leave the network scale-up data itself out of the discussion for the remainder of this section and instead try to evaluate the influence of the multiplier data on estimating this parameter.

The SMT subgroup has the largest retrospective value by a very large margin. The SMT subgroup also has the largest prospective value due in a large part to having a data for a large number of cities in multiple years that also have other data sources to calibrate against (this gives a better estimate of the subgroup specific bias). Moreover, The SMT subgroup has a somewhat surprising influence with an expected value of information ratio of $1.76$ and only $16\%$ of the the SMT data sets predicted from the posterior distribution conditional on the remaining data having more influence on the estimate. This element surprise of surprise is unrelated to the effect that the SMT has on the mean multiplier proportion estimate bias as the SMT data had an unsurprising influence on that parameter. Instead this surprise seems to be related to the effect that the SMT data has on the estimate for the prevalence of injection drug use. 

The NGO subgroup is the only other subgroup that stands out in this instance. While this subgroup doesn't have a very large effect on the estimate of $\mu$ on the whole, its effect is almost $3$ times what would be expected according to the rest of the model. As with the SMT subgroup data, the NGO subgroup also has a surprising effect on the prevalence estimates that leads to the influence that it has on the network scale-up bias as will be seen in the next section. Unlike the SMT subgroup there is a clear feature of the NGO subgroup data that may point to a violation of a model assumption. In particular, the average proportion of injection drug users that are registered to an NGO in a city is estimated to be $.27$ in 2010 but is $.51$ in 2015. No other subgroup changes this dramatically. It is not altogether clear if the NGO subgroup is defined in exactly the same way in different years. For instance, it may be that more organizations are included in the 2015 survey than in 2010 survey. 

The average expected value of information ratio for estimating $\mu$ is $1.19$. So the average subgroup is slightly more surprising than would be expected based on the rest of the data. This is not extremely worrisome as this average is inflated by the NGO subgroup, which does not have a large impact on the estimated value for $\mu$. The other subgroups yield about as consistent an estimate of $\mu$ as we would expect based on the model. 

\subsection{Evaluating the Influence of Each Site}
\label{sec:siteInf}

\begin{figure}[!h]
	\centering
	\includegraphics[width=7cm]{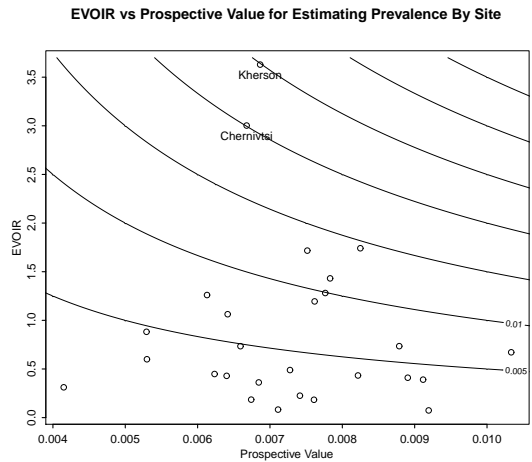}
	\caption{The expected value of information ratio plotted against the prospective expected value of information with contours for the retrospective value of information for the same partition of data.}
	\label{fig:site}
\end{figure}

In this section we will perform a retrospective value of information analysis to evaluate the contribution of each site to estimating the prevalence of injection drug use in Ukraine.  It can be seen from Figure \ref{fig:site} that the sites have very similar prospective values reflecting the fact that each of the sites tends to appear at about the same frequency and in mostly the same manner. This similarity results in the contours that indicate retrospective value being more horizontally oriented indicating that higher levels of influence are due largely to higher levels of surprise.

The majority of the sites are fairly consistent with each other having expected value of information ratios within a typical range. Two sites stand out having influenced the prevalence estimates 3 times more than what would have been expected based on the remaining data: Kherson and Chernivtsi. Neither city is of an unusual size compared to the rest of the sample, nor do they have prevalence of an unusual magnitude. Both sites do exhibit prevalence estimate spikes in 2013 that disappear in the following year, the year for which these sites have the highest expected value of information ratio. This suggests that the Survey subgroup is giving results for these two sites in that year that are incompatible with the estimates based on the rest of the data. This casts some doubts on the estimates for these sites during this year.

In this case it is interesting to note something that we do not see. In particular, the sites in the Autonomous Republic of Crimea do not seem to behave differently than the other sites for which there are data. 

\subsection{Evaluating Future Observations}
\label{sec:futObs}

We will begin this section by evaluating the prospective expected value associated with collecting more data for each multiplier subgroup, the network-scale up data, and a new multiplier data subgroup in a future year. We again consider the effect on estimating the prevalence of injection drug use using a squared error loss function.

\begin{figure}[!h]
	\centering
	\includegraphics[width=7cm]{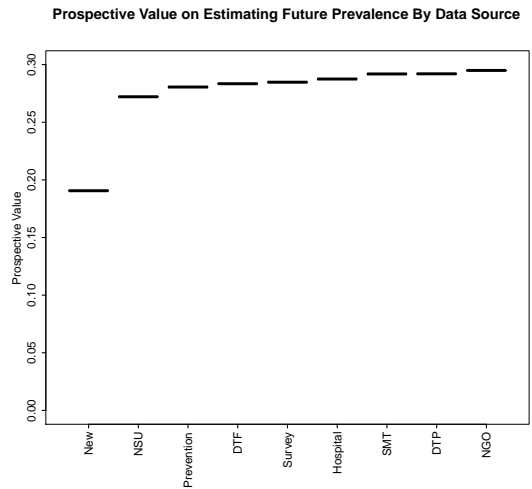}
	\caption{The prospective expected value of information for estimating the prevalence of injection drug use associated with collecting data for each multiplier subgroup, a new subgroup, and a network scale-up survey at each site in the sample during a future year. }
	\label{fig:prospective}
\end{figure}

Figure \ref{fig:prospective} shows the expected prospective value of sample information for gathering a sample for the 27 sites under consideration in a future year. While all of the data sources provide a substantial improvement in the estimation, gathering data for a new subgroup has the lowest prospective value. This is because the new subgroup would never appear in the same year as another data source (something which would provide more information about subgroup specific bias in estimation). We would then be left with having to guess the bias for the new subgroup using the general range of subgroup specific bias as estimated from the other groups and the prevalence of injection drug users in previous years. This further reinforces the need to collect multiple years of each data source
. 

The subgroups which provide the most information about the prevalence of injection drug use in future years are those years for which we already have multiple observations for. It can be seen, however, that there is a diminishing return on having multiple years of data. Collecting more NGO is deemed most valuable. It would also seem worthwhile to collect more data from the NGO subgroup to have more information about the inconsistencies with the data source discussed previously.

\section{Conclusion}

In the preceding sections we have seen that while the majority of the data sources are relatively consistent and compatible with the model being applied, the NGO subgroup is unusual in both its characteristics and influence on the fit of the model. It has been speculated that this may be a result of the definition of the subgroup changing between years, but this has not been confirmed. Collecting more data on the NGO subgroup with this in mind may help shed light on the explanation for this phenomenon. 

For future years collecting more data on data sources we have already observed rather than new sources is of a high priority as this allows us to asses the source specific bias of the associated estimates with more precision. In order to have more robust prevalence estimates in a year it is also recommended to have to collect multiple sources of data in a year.

The general consistency of the data with the model may be interpreted as troubling to those applying the multiplier and network scale-up methods to estimate population sizes without adjusting for any sources of bias. Under the fitted model, these methods perform very badly and can result in estimates of the wrong order of magnitude. The simulation used suggests that much better results can be obtained using very similar estimation methods by correcting for this bias in a direct manner. Although it has been suggest that combining multiple sources of information may help alleviate systematic errors in estimation \citep{Abdul2015}, the fit of this model indicates that doing this without correcting for the systematic error will not help. Despite the generally poor level of performance ascribed to the two methods, the multiplier method is expected to perform better than the network scale-up method.

\cite{jackson2019} also apply a value of information framework in a similar setting. While there is some overlap in goals in this paper, the main focus of the two papers is different. \cite{jackson2019} apply a prospective analysis to evaluate how knowledge about particular parameters contributes to decisions and identify what quantity of data should be collected in the future. In this paper, we are primarily interested in evaluating how different sources of information have already influenced decisions, evaluating how consistent these sources of information are, identifying outlying portions of data that may signal violations of assumptions, and forming recommendations for what estimation methods and practices should be used in the future. Our example code is now available at https://github.com/JacobLParsons/Value-of-Information.

\section{Acknowledgment}
This research was supported by NIH/NIAID 5-R01-AI136664.

\bibliographystyle{chicago}
\bibliography{Bibliography-MM-MC}
\newpage


\end{document}